\documentclass[twocolumn,aps,prc]{revtex4}
\usepackage{pdfpages}
\usepackage{graphics}
\begin{document}

\title{\Large\bf Comment on ``Spreading widths of giant resonances in 
spherical nuclei: damped transient response" by Severyukhin et al. 
[arXiv:1703.05710].}

\author{V.Yu.~Ponomarev}

\affiliation{Institute of Nuclear Physics, 
Technical University of Darmstadt, 64289 Darmstadt, Germany}

\begin{abstract}\hspace*{-4mm}
We argue whether physics of universal approach of Severyukhin et al. 
[arXiv:1703.05710]
is approved.
\end{abstract}


\date{March 29, 2017}

\maketitle

A universal approach (UA) to describe spreading width of giant resonances in
atomic nuclei has been offered recently in Ref.~\cite{Sev}.  
We discuss below its physical content.

One reads in Summary that the authors ``suggest the way to describe spreading
widths of GRs by including the coupling between one-phonon and two-phonon 
states."\cite{Sev}. This idea already belongs to well-established 
knowledge; it is employed by many nuclear models during almost half a century. 
Accordingly, the 
above-mentioned suggestion does not sound timely as original one. 

\begin{figure}[bh]
\vspace{10mm}
\includegraphics[width=7.0cm]{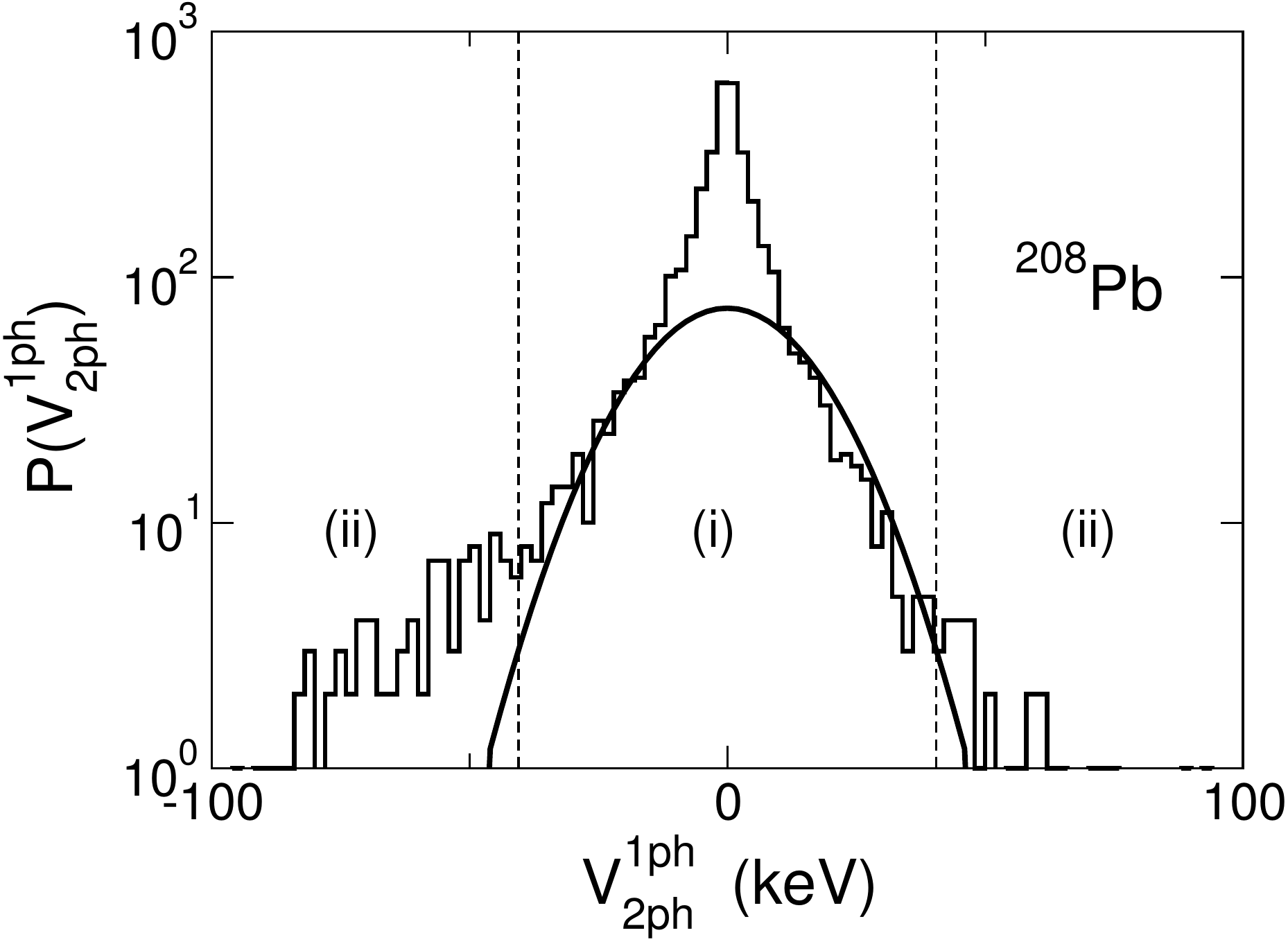}
\caption{\label{fig1}
Distribution of coupling matrix elements V$^{\rm 1ph}_{\rm 2ph}$
between the one- and two-phonon configurations in the QPM
calculation for the isoscalar giant quadrupole resonance 
in $^{208}$Pb. The solid
line denotes a Gaussian distribution expected for a chaotic
system with a width in accordance with the QPM results.
Taken from Ref.~\cite{Shev}.}
\end{figure}

An original suggestion of Severyukhin et al. is to gene\-rate the coupling 
matrix elements 
V$^{\rm 1ph}_{\rm 2ph}$
between one-pho\-non (1ph) and two-phonon (2ph) states 
``by means of the random distribution"\cite{Sev} in Gaussian form.

The matrix elements V$^{\rm 1ph}_{\rm 2ph}$
have been already analysed, e.g., 
in Refs.~\cite{Shev,Shev2} within the quasiparticle phonon model (QPM)
\cite{Sol}: They have been ``divided into two subspaces: (i) a large 
subspace with 
V$^{\rm 1ph}_{\rm 2ph}$ following the Gaussian~dis\-tribution
(plus overshoot small matrix elements) and (ii) a small subspace with large 
V$^{\rm 1ph}_{\rm 2ph}$ values above the Gaussian tails."\cite{Shev} (see
Fig.~\ref{fig1} taken from Ref.~\cite{Shev}). 
It has been demonstrated that 
``the fragmentation is dominated by the collective mechanism"\cite{Shev}, 
i.e. determined by the mat\-rix elements from the group (ii).

The UA suggests to neglect the most important matrix 
elements from the group (ii) in favour 
of less important ones from the group
(i) (without overshooting small matrix elements). 
As a result, 
calculations in Ref.~\cite{Sev} confirm\cite{com} observation 
in Ref.~\cite{Shev} 
and at the same time, are in obvious
conflict with conclusion in Ref.~\cite{Sev} that the UA ``enables
to describe gross structure of the spreading widths of the considered giant
resonances."\cite{Sev}.

To conclude: Any interaction between doorway and background states 
yields fragmentation pattern; any distribution has its width.
But this alone is not sufficient to claim that the width predicted by the UA
describes the physical width of giant resonances.
The UA appears to miss the main contribution to the width formation.

Support by the DFG (Contract No. SFB 1245) is acknowledged.\cite{com3}

\end{document}